\documentclass[12pt, amsmath, amssymb]{iopart}

\usepackage[dvips]{graphicx}
\usepackage{mathrsfs}
\usepackage{graphicx}
\usepackage{amssymb}
\usepackage{epsf,latexsym}
\usepackage{times}
\usepackage{ifpdf}
\ifpdf
\usepackage{epstopdf}   
\fi

\newcommand{\bra}[1]{\langle#1|}
\newcommand{\ket}[1]{|#1\rangle}

\newcommand{\kb}[2]{\ensuremath{\vert #1 \rangle \langle #2 \vert}}

\newcommand{\sz}[0]{\ensuremath{\mathbf{\sigma}_z}}

\renewcommand{\sp}[0]{\ensuremath{\mathbf{\sigma}_{+}}}
\newcommand{\sm}[0]{\ensuremath{\mathbf{\sigma}_{-}}}

\renewcommand{\vec}[1]{\ensuremath{{\mathrm{#1}}}}

\newcommand{\be}{\begin{equation}}
\newcommand{\ee}{\end{equation}}
\newcommand{\bea}{\begin{eqnarray}}
\newcommand{\eea}{\end{eqnarray}}

\usepackage{color}

\begin{document}

\bibliographystyle{unsrt}

\title{Strategies for Entangling Remote Spins with Unequal Coupling to an Optically Active Mediator}

\author{Erik M. Gauger}
\ead{erik.gauger@materials.ox.ac.uk}
\address{Department of Materials, University of Oxford, Parks Road, Oxford, OX1 3PH, UK}

\author{Peter P. Rohde}
\ead{peter.rohde@materials.ox.ac.uk}
\address{Department of Materials, University of Oxford, Parks Road, Oxford, OX1 3PH, UK}

\author{A. Marshall Stoneham}
\ead{a.stoneham@ucl.ac.uk}
\address{Department of Physics and Astronomy, University College London, Gower Street, London WC1E 6BT}

\author{Brendon W. Lovett}
\ead{brendon.lovett@materials.ox.ac.uk}
\address{Department of Materials, University of Oxford, Parks Road, Oxford, OX1 3PH, UK}

\date{\today}

%\frenchspacing

% ABSTRACT
\begin{abstract}
We demonstrate that two remote qubits can be entangled through an optically active intermediary even if the coupling strengths between mediator and qubits are different. This is true for a broad class of interactions. We consider two contrasting scenarios. First, we extend the analysis of a previously studied gate operation which relies on pulsed, dynamical control of the optical state and which may be performed quickly. We show that remote spins can be entangled in this case even when the intermediary coupling strengths are unequal. Second, we propose an alternative adiabatic control procedure, and find that the system requirements become even less restrictive in this case. The scheme could be tested immediately in a range of systems including molecules, quantum dots, or defects in crystals.
\end{abstract}

\maketitle

\section{Introduction}

The control and physical representation of entanglement lie at the heart of quantum computing. This leads to several design considerations. First, it is essential to have a well defined localized information carrier that can be manipulated precisely. The basic information unit is usually a qubit -- a two level quantum system. Second, it is necessary to design quantum gates to precisely control both the individual dynamics of each qubit, and their correlated motion. These two classes of operation are usually treated separately, and define single and two qubit gates. Together they form a universal set that can encode any quantum algorithm. 
Any practical quantum computer must, however, provide significantly more than just a universal gate set. It must be capable of preparing an initial quantum state, and of reading a final state. It should be able to control sufficiently many qubits for a time long enough to perform a useful calculation, without significant loss of entanglement during the evolution of the quantum state. And, as any quantum device will surely be controlled by a classical computer system, compatibility with current technologies is desirable. 

These requirements have led to a wealth of research on solid state implementations of quantum computers. In particular, an electron or nucleus with spin 1/2 constitutes a perfectly defined qubit. However, in solid state systems there is often uncertainty about the positions of and interactions between such spins, which of course makes control difficult. In this paper we shall demonstrate that it is nonetheless perfectly possible to devise very accurate quantum gates even when the parameters describing the qubits have a degree of randomness.

Many papers have discussed how spins could be used as qubits in various materials~\cite{loss98, kane98}. It is often the case that spins with the long coherence times that are so desirable for storing quantum information, are hard to control directly. However, they can be accessed indirectly through other states with shorter lifetimes, and manipulated more quickly. For example, different defects in semiconductors can be chosen such that have the correct properties play different roles: A defect with an electron spin in a quiet environment can be used to represent that quantum information, and interactions can be provided by optically active defects with shorter decoherence times~\cite{stoneham03}. Alternatively, excitons in quantum dot systems can be excited to provide coupling between dot based spin qubits~\cite{calarco03,nazir04,piermarocchi02}, that can sometimes be mediated by an optical cavity~\cite{imamoglu99a, barrett05a}. Further, in NMR quantum computing, the interaction between nuclear spins is provided by the electrons~\cite{nielsen00, vandersypen01}. A summary of such approaches can be found in Table 1.

More recently, several key experiments have been performed that demonstrate many of the ingredients that are needed for the operation of these control schemes. For example, the spin of an NV- centre in diamond can be initialized, read and manipulated optically~\cite{jelezko06}, and can be coupled to other nearby electron spins coherently, and this coherence can be manipulated optically~\cite{hanson06, gaebel06}. 
The motion of carbon-13 nuclear spins can also  be detected optically in the NV- system~\cite{childress06}, and it has even been possible to map the NV- electron spin state onto a nearby C-13 nucleus, and get it back again~\cite{dutt07}.
In semiconductor quantum dots the coherence of electron spins  has been optically controlled~\cite{greilich06} and initialized~\cite{xu07} and tunnel coupling between two electron spins in neighbouring quantum dots has been detected optically~\cite{robledo08}. Remote spin coupling through a spin bath has been demonstrated in a lithographic quantum dot system~\cite{craig04}. Electrons spin states have been used to manipulate nuclear spin qubits in other systems as well~\cite{morton06, hodges07}, in a time shorter than needed for direct addressing.

\begin{table}
\begin{center}
\begin{tabular}{|c|c|c|c|}
\hline
Qubit & Control & Method & Reference \\ \hline
Electron donor spin & Electron donor spin & Optical excitation & Ref. \cite{stoneham03} \\
QD electron spin & Exciton & Optical excitation  & Ref. \cite{calarco03,nazir04,piermarocchi02}  \\
Nuclear donor spin  & Electron spin & Electric Field & Ref. \cite{kane98} \\
Molecular nuclear spin & Electron spin & RF pulse & Refs. \cite{nielsen00, vandersypen01} \\
\hline
\end{tabular}
\end{center}
\label{states}
\caption{Summary of proposals for generating pairwise entanglement of qubits with the help of controlled intermediaries. Different approaches will be optimized by choosing different material systems.}
\end{table}

In this paper, we consider explictly a system for entangling two such long lived spin qubits, via a third, central, electron spin of a different species that might have a much shorter decoherence time. This central spin can be optically excited, and in the excited state the electron wavefunction typically has a greater spatial extent than in the ground state (see Fig. 1). If this larger wavefunction overlaps with the two neighbouring spins, it gives rise to an exchange coupling. This possibility was first raised by Stoneham {\it et al.}~\cite{stoneham03} who introduced a scheme for entangling deep donor spins in silicon, where various defects could be used  for the different spin species (for example. Mg$^+$, Se$^+$, or Bi$^+$\cite{stoneham03, stoneham05} are good candidates).
In contrast to the previous work, we shall here study a more general situation in which there is only limited control over the various coupling parameters in the problem, and show that it is still possible to obtain a highly entangling gate operation.

\section{Model}

Let us first introduce a general model for the system we are considering. It has the following Hamiltonian, in the usual notation using Pauli spin operators $\sigma$\footnote{We use $\sigma$ matrices that all have eigenvalues $\pm 1$.}:
\bea
H  &=&  E_Q \sz^Q + E_C \sz^C + E_{Q'} \sz^{Q'} +  \nonumber \\
&&\ket{e} \left( J_1 (\vec{\sigma}^Q \cdot  \vec{\sigma}^C - \alpha \vec{\sigma}_z^Q \cdot \vec{\sigma}_z^C)  +  J_2 (\vec{\sigma}^{Q'} \cdot \vec{\sigma}^C  - \alpha \vec{\sigma}_z^{Q'} \cdot  \vec{\sigma}_z^C) + \omega_0  \right)  \bra{e},
\label{eq:hamiltonian}
\eea
 $Q$ and $Q'$ label the two qubit spins; $C$ is the central (control) spin which has two degrees of freedom: one is its spin $\sigma^C$ and the other is its orbital state which we restrict to the space spanned by states $\{\ket{g}, \ket{e}\}$. We assume an allowed optical transition of energy $\omega_0$ between $\ket{g}$ and $\ket{e}$, but that $Q$ and $Q'$ do not couple directly to an optical field.
 Each of the $E_j$ gives the Zeeman splitting of spin $j$ in an external magnetic field of strength $B$ ($E_j  = \mu_j B$). $J_1$, $J_2$ is the exchange coupling between spins $Q$ or $Q'$ and $C$ respectively, which is only present when $C$ is in the excited state $\ket{e}$. For $\alpha=0$ it takes an isotropic Heisenberg form, and for $\alpha = 1$ it represents an $XY$ type coupling. We assume that the control-qubit coupling strength when the control is in the $\ket{g}$ state is negligible in comparison with the coupling when the control is in the $\ket{e}$ state. Calculations in Ref.~\cite{wu07} show that this is indeed the case for donors in silicon, where the $\ket{e}$ coupling can be two to four orders of magnitude larger. 
We also ignore direct donor-donor coupling, which again is valid for the silicon donor system.  
For suitable spatial configuration with donors separated by around 25~nm, direct donor - donor coupling can be as low as $4.73 \times 10^{-3}~\mathrm{GHz}$ whereas the control mediated coupling strength is still up to  $157~\mathrm{GHz}$ \cite{kerridge06, kerridge07}.

In order to control the interaction, a laser is applied with frequency $\omega_l$, and so introduces an oscillatory term into the Hamiltonian. The oscillation can be removed by transforming into a frame rotating at $\omega_l$ and making a rotating wave approximation, whereupon we write:
\bea
H  &=&  E_Q \sz^Q + E_C \sz^C + E_{Q'} \sz^{Q'} +  \frac{\Omega(t)}{2} \left( \kb{e}{g} + \kb{g}{e} \right) \nonumber \\
&&+\ket{e} \left( J_1 (\vec{\sigma}^Q \cdot  \vec{\sigma}^C - \alpha \vec{\sigma}_z^Q \cdot \vec{\sigma}_z^C)  +  J_2 (\vec{\sigma}^{Q'} \cdot \vec{\sigma}^C  - \alpha \vec{\sigma}_z^{Q'} \cdot  \vec{\sigma}_z^C) + \Delta \right)  \bra{e},
\label{eq:hamiltonian_rwa}
\eea
where $\Delta \equiv \omega_0-\omega_l$ is the laser detuning from the transition and $\Omega(t)$ is (generally time-dependent) Rabi frequency.

A general state of the three spin system $\ket{\phi}$ is given by a superposition of the spin basis states \mbox{$\ket{\{\uparrow,  \downarrow\}}_{Q} \otimes  \ket{\{\uparrow,  \downarrow\}}_{C} \otimes  \ket{\{\uparrow,  \downarrow\}}_{Q'}$}, where the arrows represent the spin up or down projection along the $z$-quantization axis. For convenience, we adopt the usual qubit notation $\ket{QCQ'}$ with $Q, C, Q'$ being either $0$ for  the `down' or $1$ for the `up' projection of the respective spin qubit. 

\begin{figure}
\begin{center}
\includegraphics[width=0.4\columnwidth]{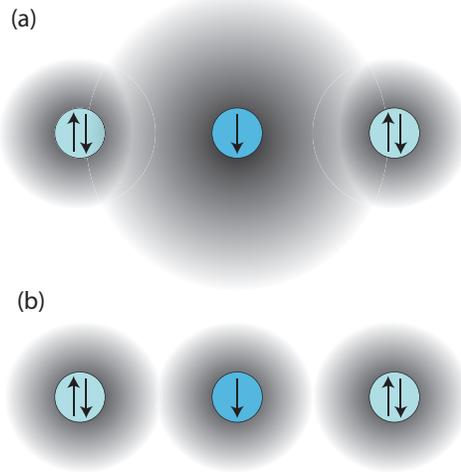}
\caption{In the optical ground state of the central control qubit (b), the wavefunctions of the three species do not overlap and there is no spin-spin exchange interaction. When the central control is optically excited (a) its wavefunction has a larger extent and so activates the spin-spin coupling.}
\end{center}
\end{figure}

We shall be concerned with the situation in which we initialize the system in $\ket{g}$ and where qubit $C$ is prepared in the state $\ket{0}$. We then allow the system to evolve under laser excitation until there is a `revival' such that $C$ returns to the state $\ket{0}$. 
We will show that the remaining two qubits $Q$ and $Q'$ can become entangled by such an operation. In the following, we will explore two contrasting scenarios, First, a fast `dynamic' optical excitation in which the system is excited suddenly by a pulsed laser, then allowed to evolve for a time before sudden de-excitation. Second, we will look at an `adiabatic' approach in which the laser intensity and/or frequency is changed slowly and continuously such that the system follows its instantaneous eigenstates.

\section{Dynamic Excitation} \label{sec:dyn_excitation}

Consider first a laser that is resonant with the $\ket{g} - \ket{e}$ transition (i.e. $\Delta = 0$). If a rectangular pulse is applied for a time $t_l = \pi/\Omega$ all population is transferred from $\ket{g}$ to $\ket{e}$ thus activating the spin couplings. After waiting for a specified amount of time this interaction is deactivated again using an identical pulse.
For the dynamical approach to work, we require a system where $\alpha=1$, i.e. that the Ising part is removed from the Heisenberg interaction and we are left with an $XY$ coupling. We will discuss the reason for this at the end of this section.
Hamiltonians not satisfying $\alpha=1$ are not amenable to the dynamic method and the more general adiabatic approach discuss later must be used. 

Let us assume that the optical excitation is fast in comparison to the subsequent spin dynamics, which are described by the restricted Hamiltonian $\bra{e}H\ket{e}$:
\be
\hspace{-2cm}H_e  =  E_Q \sz^Q + E_C \sz^C + E_{Q'} \sz^{Q'} +  \left( J_1 (\vec{\sigma}_x^Q \cdot  \vec{\sigma}_x^C +\vec{\sigma}_y^Q \cdot \vec{\sigma}_y^C)  +  J_2 (\vec{\sigma}^{Q'}_x \cdot \vec{\sigma}^C_x  +\vec{\sigma}_y^{Q'} \cdot  \vec{\sigma}_y^C) \right).
\ee
$H_e$ conserves the total spin projection: $\Sigma_z = \sigma_z^Q + \sigma_z^C + \sigma_z^{Q'}$. Therefore the evolution can be partitioned into subspaces of different $\Sigma_z$:
\be
H_e = H_0 \oplus H_1 \oplus H_2 \oplus H_3
\ee
where $H_i$ is the Hamiltonian of the subspace with $\Sigma_z = 2i-3$ (see Table 2). The central (control) qubit is set to $\ket{0}$ initially and so we need not consider the $H_3$ space further. Let us now analyze the dynamics of the other subspaces.

\begin{table}
\begin{center}
\begin{tabular}{|c|c|c|}
\hline
Subspace & $\Sigma_z$ & Component States \\ \hline
$H_3$ & 3 & $\ket{111}$\\
$H_2$ & 1 & $\{\ket{110}, \ket{101}, \ket{011}\}$\\
$H_1$ & -1 & $\{\ket{100}, \ket{010}, \ket{001}\}$\\
$H_0$ & -3& $\ket{000}$\\
\hline
\end{tabular}
\end{center}
\label{states}
\caption{Table showing the four uncoupled subspace and the notation used for each.}
\end{table}

\subsection{$H_1$ subspace ($\Sigma_z = -1$)}
In the basis of states $\{\ket{010}, \ket{100}, \ket{001}\}$, and for qubits with the same $g$-factors ($E_Q = E_{Q'}$) we have
\be
H_1 = E_C \left(\begin{array}{ccc}
R & J'_1 & J'_2 \\
J'_1 & 1 & 0 \\
J'_2 & 0 & 1
\end{array} \right),
\ee
with $R \equiv (2E_Q/E_C)-1$, $J'_1 \equiv  2 J_1/E_C$ and $J'_2 \equiv 2 J_2/E_C$. There is always one eigenvector that is orthogonal to $\ket{A}\equiv\ket{010}$:
\be
\ket{E} = \frac{J'_1\ket{100} -J'_2 \ket{001}}{\sqrt{{J'_1}^2+{J'_2}^2}}.
\ee
Let us define a state that is orthogonal to both $\ket{A}$ and $\ket{E}$:
\be
\ket{T} = \frac{J'_2\ket{100} +J'_1 \ket{001}}{\sqrt{{J'_1}^2+{J'_2}^2}},
\ee
and rewrite the Hamiltonian in the basis $\{\ket{A}, \ket{T}, \ket{E}\}$
\be
H_1 = E_C\left(\begin{array}{cc|c}
R & \sqrt{{J'_1}^2 + {J'_2}^2} & 0 \\
\sqrt{{J'_1}^2 + {J'_2}^2} & 1 & 0 \\
\hline
0 & 0 & 1
\end{array} \right).
\ee
Prior to laser excitation, the system contains no component of $\ket{A}$; it can therefore be written as a superposition of $\ket{E}$ and $\ket{T}$. $\ket{E}$ is an eigenstate so only accumulates phase. $\ket{T}$ undergoes Rabi cycling to $\ket{A}$; after each cycle all population returns to $\ket{T}$, and the system `revives' such that no excitation is left on the central qubit. This `revival time' $t_\mathrm{rev}$ is given by:
\be
t_\mathrm{rev} = \frac{2 n \pi}{E_C\sqrt{(R-1)^2 +4{J'_1}^2 + 4{J'_2}^2}},
\label{eq:rev_time}
\ee
where $n$ is some integer, the number of oscillations that have occurred. At revival, we have 
\begin{eqnarray}
\ket{T} &\to& e^{i\theta_T} \ket{T},\nonumber\\
\ket{E} &\to& e^{i\theta_E} \ket{E}
\end{eqnarray}
where
\begin{eqnarray}
\theta_T &=& \pi n-\frac{E_C(1+R)t_\mathrm{rev}}{2} = \pi n\left (1- \frac{(R+1)}{\sqrt{(R-1)^2 +4{J'_1}^2 + 4{J'_2}^2}}\right),\nonumber\\
\theta_E &=& -E_C t_\mathrm{rev} = - \frac{2 n \pi}{\sqrt{(R-1)^2 +4{J'_1}^2 + 4{J'_2}^2}}.
\end{eqnarray}

\subsection{$H_2$ subspace ($\Sigma_z = 1$)}

In the basis $\{\ket{101},\ket{011},\ket{110}\}$ we can write
\be
H_2 = E_C\left(\begin{array}{ccc}
-R & J'_1 & J'_2 \\
J'_1 & -1 & 0 \\
J'_2 & 0 & -1
\end{array} \right).
\ee
After the same revival time $t_{\rm rev}$, $C$ again returns to the $\ket{0}$ state, such that the state $\ket{A'}\equiv\ket{101}$ undergoes the following transformation:
\be
\ket{A'} \to e^{ i \theta_{A'}}\ket{A'},
\end{equation}
where
\begin{equation}
\theta_{A'} = \pi n+\frac{E_C(1+R)t_\mathrm{rev}}{2} = \pi n\left (1+ \frac{(R+1)}{\sqrt{(R-1)^2 +4{J'_1}^2 + 4{J'_2}^2}}\right) .\nonumber\\
\end{equation}

\subsection{$H_0$ subspace ($\Sigma_z = -3$)}

Finally, we have
\begin{equation}
\ket{000} \to e^{i\theta_Z}\ket{000},
\end{equation}
where
\begin{equation}
\theta_Z = -E_C(R+2)t_\mathrm{rev} = -\frac{2(R+2) n \pi}{\sqrt{(R-1)^2 +4{J'_1}^2 + 4{J'_2}^2}}.
\end{equation}

\subsection{Evolution of the logical qubits; entangling power}

Combining the dynamics for the different subspaces, the overall unitary evolution of the system in the logical basis of qubits $Q$ and $Q'$ ($\{\ket{00},\ket{01},\ket{10},\ket{11}\}$) is
\begin{equation}
U' = \left(\begin{array}{cccc}
e^{i\theta_Z} & 0 & 0 & 0 \\
0 & \Delta_1 & \Delta_2 & 0 \\
0 & \Delta_2 &  \Delta_3 & 0 \\
0 & 0 & 0 & e^{i\theta_{A'}},
\end{array} \right),
\label{eq:dyn_action}
\end{equation}
where
\begin{eqnarray}
\Delta_1 &=& \frac{e^{i\theta_E}{J'_1}^2 + e^{i \theta_T} {J'_2}^2}{{J'_1}^2 + {J'_2}^2}, \nonumber\\
\Delta_2 &=& \frac{J'_1 J'_2 \left(e^{i \theta_T} - e^{i \theta_E}\right)}{{J'_1}^2 + {J'_2}^2} \nonumber\\
\Delta_3 &=& \frac{e^{i\theta_T}{J'_1}^2 + e^{i \theta_E} {J'_2}^2}{{J'_1}^2 + {J'_2}^2}.
\end{eqnarray}

To determine the extent to which this evolution creates entanglement, 
we use the measure of average gate entangling power developed by Zanardi, {\it et al.}~\cite{zanardi00}, who considered a bipartite state $\ket{\Psi}$, which lives in a Hilbert space ${\cal H}_1 \otimes {\cal H}_2$.  The entangling power of $U$ is found by taking the average of the linear entropy of the reduced density matrix ($\rho_1 = \tr_1[\ket{\Psi}]$) over a uniform distribution of input product states $\ket{\psi_1} \otimes \ket{\psi_2}$: 
\begin{equation} \label{eU}
e(U) = \overline{E(U\ket{\psi_1}\otimes\ket{\psi_2})}^{\psi_1,\psi_2},
\end{equation}
$E(\ket\Psi) = 1 - \mathrm{tr}({\rho_1}^2)$ is the linear entropy of $\rho_1$. The maximum value of the entangling power is about 0.22, and it falls to zero for a gate that produces no entanglement. 
Using Eq. 5 from \cite{zanardi00} we determine that  
\begin{eqnarray}
e(U) &=& \frac{1}{18}(8 - 2|\Delta_1|^2 - |\Delta_1|^4 - 4|\Delta_2|^2 - 2|\Delta_2|^4\nonumber\\
&-& 2|\Delta_3|^2 - |\Delta_3|^4 - 2\mathrm{Re}[e^{i(\theta_Z + \theta_{A'})} {\Delta_2}^2]\nonumber\\
&-& 2\mathrm{Re}[e^{i(\theta_Z + \theta_{A'})} {\Delta_1} {\Delta_3}]).
\label{eqn:measure_entangling_power}
\end{eqnarray}

\begin{figure}
\begin{center}
\includegraphics[width=0.5\columnwidth]{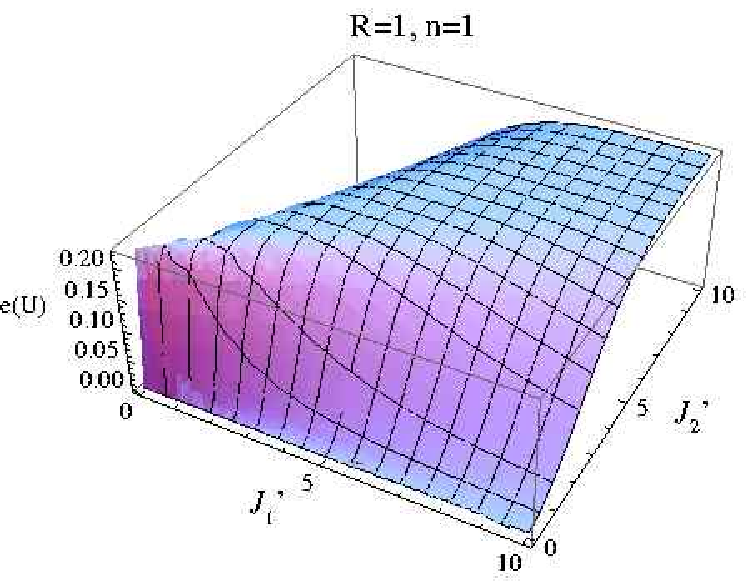} \hspace{1cm}
\includegraphics[width=0.4\columnwidth]{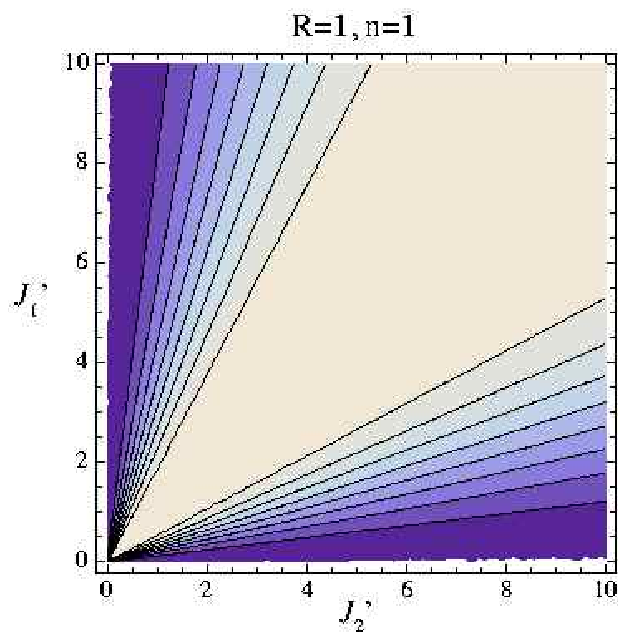}
\caption{Average entangling power $e(U)$ of the dynamic two qubit gate after the first revival $n=1$.} 
\label{dynamic_eU_plots}
\end{center}
\end{figure}

\begin{figure}
\begin{center}
\includegraphics[width=0.6\columnwidth]{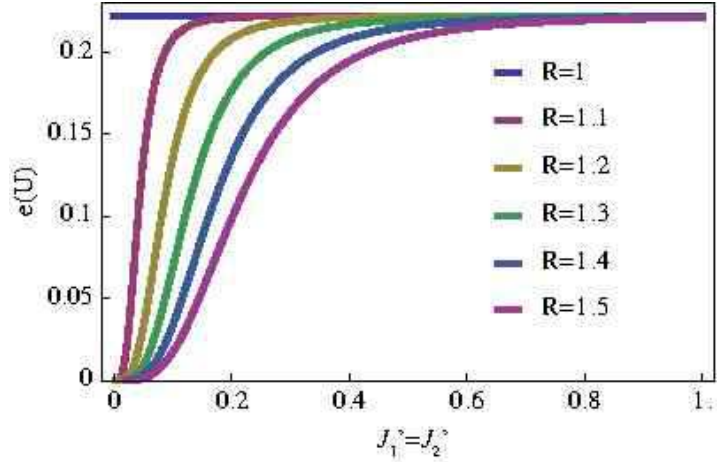}
\caption{Average entangling power of the dynamic gate when $J_1=J_2$, for different values of $R \equiv (2E_Q/E_C)-1$.} \label{dynamic_eU_j1j2_plot}
\end{center}
\end{figure}

Fig. \ref{dynamic_eU_plots} shows the average entangling power of the gate, $e(U)$, after the first revival $n=1$, for $R=1$. \footnote{For even $n$ the average entangling power vanishes, $e(U)=0$, whereas for odd $n$ the entanglement revives to the same level as for $n=1$.} The entangling power drops to zero when either $J_1 = 0$ or $J_2=0$: If either of the qubits is not interacting with the central control, then no entanglement is possible. By contrast $e(U)$ is maximized (and reaches its theoretical maximum) when $J_1'=J_2'$ -- and in Fig. \ref{dynamic_eU_j1j2_plot} we show $e(U)$ for different values of the ratio $R$ in this equal coupling case. As $R$ get closer to unity, the entangling power is larger for  smaller $J_1'=J_2'$. For larger $J_1'=J_2'$ all plots approach maximal entangling power. Overall, we can conclude that the dynamic gate has a reasonable entangling power over a wide range of parameter space.

We can see that it is essential that $\alpha = 1$ for the dynamic approach to work; if this were not the case then the revival times in the $H_1$ and $H_2$ subspaces would not coincide, invalidating the analysis presented here. In order to overcome this restriction we must change our strategy and we shall discuss this next.

\section{Adiabatic Excitation}

An alternative method for creating entanglement in our system relies on adiabatic following of eigenstates. It can be implemented by slowly modulating the intensity of a laser that is close to resonance with the optical transition of the central qubit. Prior to excitation, the system is prepared in a superposition of the computational basis states, $\ket{QQ'} \in \{\ket{00}, \ket{01}, \ket{10}, \ket{11}\}$. The laser intensity is then varied such that adiabatic following of eigenstates occurs, so if the intensity is decreased again population returns to the computational basis.

\begin{figure}
\begin{center}
\includegraphics[width=0.6\columnwidth]{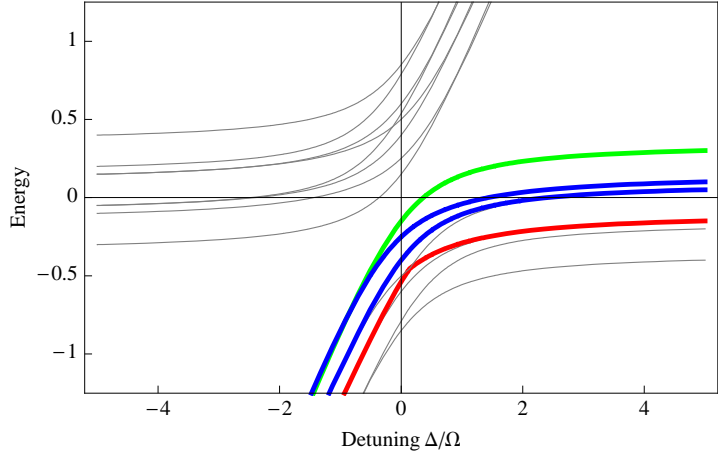}
\caption{Example of eigenspectrum as a function of the detuning $\Delta$  for typical parameters. The eigenstates tending to the computational basis states for $\Omega / \Delta \to 0$  (i.~e. to the far right in this figure) are colour-coded as follows: $\ket{00}$: green, $\ket{10}$ and $\ket{01}$: blue, $\ket{11}$: red. }
\label{fig:eigenspectrum}
\end{center}
\end{figure}

With the laser on, each of the eigenstates consists of some superposition of $\ket{g}$ and interacting $\ket{e}$ levels, such that the eigenenergies are determined not only by the optical coupling but also by the Heisenberg interaction between the spins. Fig.~\ref{fig:eigenspectrum} shows such an eigenspectrum of Hamiltonian (\ref{eq:hamiltonian_rwa}) as a function of $\Delta / \Omega$.  For $ \Delta/\Omega  \to \infty $, the eigenenergies tend to the Zeeman split levels comprising the logical basis. 
The relative spacing between the eigenstates changes with laser intensity - i.e. when $\Delta / \Omega$ approaches zero. Each eigenstate therefore acquires a different dynamical phase as a consequence of its time evolution, which results in different final phases of the logical states -- and thus enables the implementation of a controlled phase gate.

In order to determine which pulse shape and temporal profile is suitable for achieving adiabatic following, we shall for the moment neglect the coupling between the spins in the excited levels. This gives us eight uncoupled two level systems (2LS), each of which is driven independently by the laser. In this case it is straightforward to derive an `adiabaticity condition' which ensures eigenstate following, with suppressed transitions between the eigenstates:
\begin{equation}
\frac{ \dot{\Omega}(t)\Delta(t) - \Omega(t) \dot{\Delta}(t)}{2 [\Delta(t)^2 + \Omega(t)^2]^{3/2}} \ll 1,
\label{eqn:adiabaticity_condition}
\end{equation}
as in Landau-Zener theory. This condition can be derived by analysing the time-dependent unitary transformation of a driven 2LS Hamiltonian in the diagonal basis and stipulating that the coupling between eigenstates should be small compared to their energetic spacing \cite{gauger07c}. For a Gaussian profile of the laser intensity,
\begin{equation}
\Omega(t) = \Omega_0 \exp[-(t/\tau)^2],
\label{eq:gaussian_pulse}
\end{equation}
and constant detuning  $\Delta$, inequality (\ref{eqn:adiabaticity_condition}) can be satisfied by demanding $\Omega_0 / \Delta^2 \ll \tau$. Adiabatic following is therefore always achieved in the limit $\Omega_0 \ll \Delta$ together with a sufficiently large pulse duration $\tau$.  \footnote{For small $\Omega_0 / \Delta $, $\tau$ must be automatically large since the interacting levels are only weakly excited.}

The spin-spin interactions mean that the system cannot be regarded as eight separate 2LS. Rather, the eigenstates are coupled states which ultimately generate the desired entanglement. Nonetheless, the inequality  (\ref{eqn:adiabaticity_condition}) is still a requirement for achieving adiabatic following, but it is not always sufficient; it is also essential that the eigenstates belonging to the computational subspace without laser irradiation must be energetically distinguishable from those outside this subspace. This avoids population leakage from of the computational basis associated with a mixing of eigenstates - and is reasonable in our scheme which presupposes two different species for $Q$ and $C$.

\subsection{Action of the adiabatic operation}

As in Section \ref{sec:dyn_excitation}, it suffices to analyze different $\Sigma_z$ subspaces separately. Unlike for the dynamic excitation, however, the restriction $\alpha=1$ is not a requirement for the adiabatic scheme (and neither is $E_{Q} = E_{Q'}$). Once more, the control qubit should be initialized to  $\sz = -1$ \footnote{It is important to have the central qubit in a well-defined initial state because of subspace dependent energy shifts of the eigenstates. In general, this leads to different pulse durations for a successful entangling operation depending on whether $C$ is initially in the $\sz = -1$ or the $\sz = +1$ state.}. The zero excitation subspace then only contains the logical $\ket{00}$ state. Similarly, the two excitation subspace contains only the logical $\ket{11}$ state. Therefore, no population transfer between these and other logical states is possible and each merely accumulates a phase after the adiabatic pulse has been applied. The situation is different for the single excitation subspace, which is populated by the two interacting states $\ket{01}$ and $\ket{10}$. For this subspace, a more complex unitary operation between the logical states results from the adiabatic operation.

The most general action of the adiabatic operation can be therefore described by the following unitary matrix in the basis of the logical states $\{ \ket{00, \ket{01}, \ket{10}, \ket{11}} \}$:
\begin{equation}
U_{ad} = \left(\begin{array}{cccc}
e^{i \phi_{00}} & 0 & 0 & 0 \\
0 & \psi & \chi' & 0 \\
0 &  \chi  & \psi' & 0 \\
0 & 0 & 0 & e^{i \phi_{11}} 
\end{array}\right),
\label{eqn:ad_action}
\end{equation}
where $\phi_{ij}$ is the phase acquired by the state $\ket{ij}$.  The coefficients $\psi$, $\psi'$, $\chi$ and $\chi'$ form a unitary $ 2 \times 2$ submatrix accounting for population transfer between $\ket{10}$ and $\ket{01}$ as well as the phase acquired by each of these two states. The complexity of the Hamiltonian (\ref{eq:hamiltonian_rwa}) makes it difficult to find analytical expressions for the elements of Eq. (\ref{eqn:ad_action}). Nevertheless, for a given set of system and laser control parameters $U_{ad}$ is straightforward to obtain numerically. 

The structure of $U_{ad}$ takes the same form as Eq. (\ref{eq:dyn_action}) obtained in Sec.  \ref{sec:dyn_excitation} for the dynamic operation. This enables a direct comparison of the entangling power of the dynamic and adiabatic approach using the measure defined in Eq.~(\ref{eqn:measure_entangling_power}).

\subsection{Adiabatic CPHASE Gate}

Under certain conditions the off-diagonal terms in Eq. (\ref{eqn:ad_action}) are zero. Using the numerical techniques discussed earlier, we find that this is the case where either:
\begin{enumerate}
\item there are degenerate logical qubits $E_{Q} = E_{Q'}$ with equal coupling $J_1 = J_2$ to the control qubit, or
\item there are non-degenerate logical qubits $E_{Q} \neq E_{Q'}$ and a pulse duration $\tau$ longer than $(E_{Q} - E_{Q'})^{-1}$.
\end{enumerate}
In this case the adiabatic operation is simply
\begin{equation}
U_{phase} = \left(\begin{array}{cccc}
e^{i \phi_{00}} & 0 & 0 & 0 \\
0 & e^{i \phi_{01}} & 0 & 0 \\
0 &  0  & e^{i \phi_{10}} & 0 \\
0 & 0 & 0 & e^{i \phi_{11}} 
\end{array}\right) ,
\label{eq:phase_gate}
\end{equation}
which is locally equivalent to the CPHASE gate when~\cite{calarco03, lovett05}:
\begin{equation}
\varphi = \phi_{00} - \phi_{01} - \phi_{10} + \phi_{11} = \pi .
\label{eq:non_trivial_phase}
\end{equation}
This condition that can always be satisfied by choosing an appropriate pulse duration.

\begin{figure}
\begin{center}
\includegraphics[width=0.6\columnwidth]{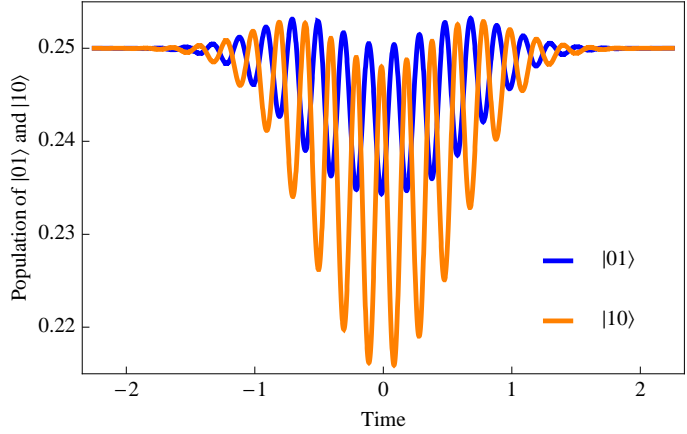}
\caption{Interference of population between $\ket{10}_L$ and $\ket{01}_L$ in the $i=1$ subspace. This effect is a consequence of the time evolution of a superposition of eigenstates as explained in the main text. As shown,  a sufficiently long pulse duration damps the oscillations out and restores all population back into the original levels at the end of the pulse. The Gaussian pulse is centred around $t=0$, where time is given in units of the pulse width $\tau$, which in this case is set to 150~ps. }
\label{fig:population_interference}
\end{center}
\end{figure}

We now explain why a system which satisfies the less restrictive set of conditions (ii) above gives a unitary operation of CPHASE form. 
Our explanation need only consider the $i=1$ subspace since the others are in CPHASE form under any conditions. We write a general initial state in this subspace characterized by the amplitudes $\alpha$ and $\beta$ as 
\be
\ket{\psi(0)} =  \alpha \ket{100} + \beta \ket{001} .
\ee
Adiabatic following of eigenstates means that this state evolves under the influence of the laser into
\be
\ket{\psi(t)} = \alpha \ket{\mu(t)} + \beta \ket{\nu(t)},
\label{eq:es_superpos}
\ee
where $\ket{\mu(t)}$ tends to $\ket{100}$ and $\ket{\nu(t)}$ to $\ket{001}$ as the Rabi frequency $\Omega(t)$ goes to zero. The time evolution of the slowly changing constituent eigenstates of Eq. (\ref{eq:es_superpos}) follows
\bea
\ket{\mu(t)} & = & e^{i E_{\mu} t} \ket{\mu}, \\
\ket{\nu(t)} & = & e^{i E_{\nu} t} \ket{\nu},
\eea
where  $\mu$ and $ \nu$ denote the instantaneous eigenstates and $E_{\mu}$ and $E_{\nu}$ are their associated eigenvalues. The time evolution of $\ket{\psi(t)}$ may thus be written as
\be
\ket{\psi(t)} =  e^{i E_{\mu} t} \left( \alpha \ket{\mu} +  e^{i (E_{\nu}  - E_{\mu}) t}  \beta \ket{\nu} \right).
\label{eqn:eigsuperpos}
\ee
The $i=1$ subspace consists of three spin states in each of the ground and excited optical states, so that $\ket{\mu}$ and $\ket{\nu}$ will be composed of up to six different states. Focussing on the physical states which correspond to the two logical states, $\ket{\mu}$ and $\ket{\nu}$ can be written as follows:
\bea
\ket{\mu} & = & (p \ket{100} + q \ket{001}) \otimes \ket{g} + \ket{m}, \label{eqn:eigmu} \\
\ket{\nu} & = & (r \ket{100} + s \ket{001}) \otimes \ket{g} +  \ket{n}, \label{eqn:eignu}
\eea
where $p, r, q$ and $s$ are appropriate amplitudes of this decomposition and $\ket{m}$ and $\ket{n}$ contain the contributions of the four remaining states $\ket{010}\ket{g}, \ket{100}\ket{e}, \ket{001}\ket{e}$  and  $\ket{010}\ket{e}$. Inserting Eqs. (\ref{eqn:eigmu}, \ref{eqn:eignu}) into Eq. (\ref{eqn:eigsuperpos}) yields
\bea
\ket{\psi}(t) & = & e^{i E_{\mu} t} \left( \alpha \ket{m} +  e^{i (E_{\nu}  - E_{\mu}) t}  \beta \ket{n} \right.  \nonumber \\ 
 & + &   (\alpha p +  e^{i (E_{\nu}  - E_{\mu}) t}  \beta r) ~ \ket{100}  \nonumber \\ 
& + & \left.   (\alpha q +  e^{i (E_{\nu}  - E_{\mu}) t}  \beta s) ~ \ket{001} ~ \right).
\label{eqn:eiginterference}
\eea
At the end of the pulse, adiabaticity ensures that the first two terms of Eq.~(\ref{eqn:eiginterference}) disappear. The third and fourth terms show
cycles of constructive and destructive  interference of the states $\ket{100}\ket{g}$ and $\ket{001}\ket{g}$. The interference oscillations range between $\alpha p \pm \beta r$ for $\ket{100}\ket{g}$ and $\alpha q \pm \beta s$ for $\ket{001}\ket{g}$ with a period of $(E_{\nu}  - E_{\mu}) / 2 \pi =  \Delta E / 2 \pi$. 
Our simulations show that if the interference period $\Delta E / 2 \pi$ is considerably faster than the pulse duration $\tau$, all population is restored to the original levels at the end of the pulse, as illustrated in the scenario of Fig. \ref{fig:population_interference} -- which of course allows the realisation of a CPHASE gate with suitable laser control parameters. Conversely, comparatively fast laser pulses generally transfer population between the two logical states, leading once more to the gate described by Eq. (\ref{eqn:ad_action}). 

\subsection{Entangling Power}

We simulate the adiabatic operation by integrating Hamiltonian Eq. (\ref{eq:hamiltonian}) with a Gaussian profile of the Rabi freguency as in Eq. (\ref{eq:gaussian_pulse}). In order to prevent Landau-Zener transitions between the eigenstates, $\tau$ needs to be suitably large. Depending on whether a pure phase gate or a more general entangling gate is desired, the effect of the operation can be obtained after the pulse has finished by extracting either a non-trivial phase as in Eq. (\ref{eq:non_trivial_phase}) or the unitary matrix Eq. (\ref{eqn:ad_action}). The average entangling power of both these quantities can then be determined using Eq. (\ref{eqn:measure_entangling_power}).

We find that a more pronounced difference between the Zeeman splittings of all three spins makes the adiabatic following more robust and permits the application of a pulse with shorter duration.  This might be achieved by using species with varying $g$ values, or through an inhomogeneous magnetic field.

\begin{figure}
\begin{center}
\includegraphics[width=0.5\columnwidth]{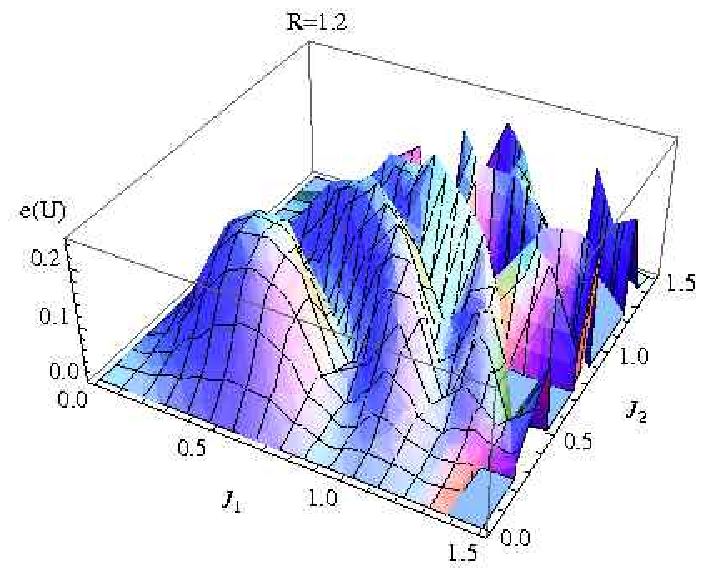} \hspace{1cm}
\includegraphics[width=0.4\columnwidth]{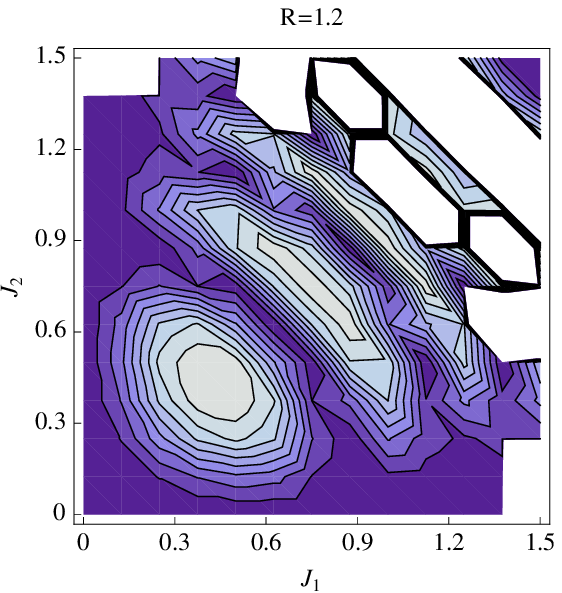}
\caption{Average entangling power $e(U)$ of the adiabatic two qubit gate. The white areas of the contour plot (light blue patches without grid lines in the 3D plot) correspond to parameter combinations for which the adiabatic gate leads to population leakage out of the computational basis, making $e(U)$ ill-defined. $J_1$ and $J_2$ are given in units of $0.1~\rm{ps}^{-1}$. The pulse duration $\tau = 0.5~\rm{ns}$, the detuning $\Delta = 0.5~\rm{ps^{-1}}$ and coupling strength $\Omega_0 = 0.3~\rm{ps^{-1}}$. N. B. For a particular  $J_1, J_2$, it is possible to optimize the speed of the gate by varying $\tau$, $\Delta$ and $\Omega$} 
\label{adiabatic_eU_plots}
\end{center}
\end{figure}

\begin{figure}
\begin{center}
\includegraphics[width=0.6\columnwidth]{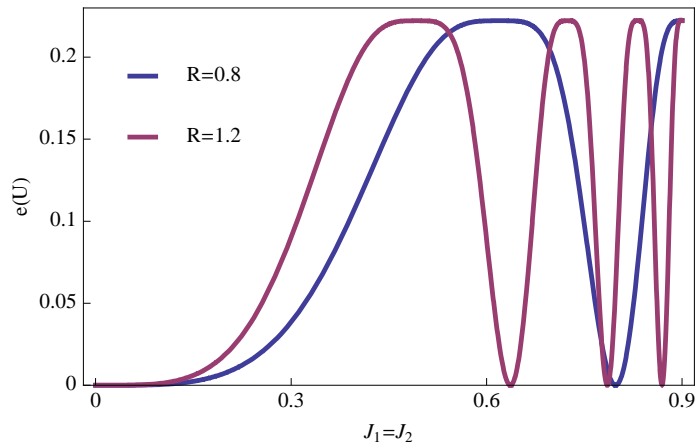}
\caption{Average entangling power of the adiabatic gate when $J_1=J_2$ (in units of $0.1~\rm{ps}^{-1}$), for different values of $R$.}
\label{adiabatic_eU_j1j2_plot}
\end{center}
\end{figure}

Fig. \ref{adiabatic_eU_plots} shows a typical plot of the average adiabatic entangling power (as in Eq. (\ref{eU})) as a function of $J_1$ and $J_2$, and Fig. \ref{adiabatic_eU_j1j2_plot} presents a cross section along the along the diagonal $J_1 = J_2$.~\footnote{In order to compare these simulations to those for the dynamic gate we choose more restrictive conditions than are necessary: Matching onsite energies $E_{Q} = E_{Q'} = 0.1~\rm{ps}^{-1}$, $XY$ type coupling ($\alpha = 1$) and $E_C = 0.1~\mathrm{ps}^{-1}$.}
As for the dynamic gate, the most entangling region occurs along the diagonal where $J_1$ and $J_2$ are equal; in contrast to the dynamic gate, the graph shows a rather complicated oscillatory structure. This is connected to how  phase accumulates in the adiabatic gate --  
which is in turn related to the values of $J_1$ and $J_2$.  The plots contain some areas where adiabatic following does not occur, causing significant population leakage away from the computational basis states. In this case $e(U)$ is ill defined.
However, a well defined adiabatic operation can always be recovered by making adjustments to the choice of laser control parameters.

\section{Decoherence}

We shall now discuss the effect of decoherence on our predictions. The dominant decoherence source or sources will be different for each physical implementation of our Hamiltonian Eq.~\ref{eq:hamiltonian}, and a full discussion of each possible process is beyond the scope of a single paper. However, we shall discuss decoherence that arises from
spontaneous decay of the optically excited state. This could either be radiative or non radiative~\cite{stonehambook, vinh08}, and would be the dominant source for deep donors in silicon~\cite{stoneham03, stoneham05}.

We use a standard quantum optical master equation \cite{cohentannoudji92} to model the decay affecting the control qubit $Q$
\begin{equation}
\dot{\rho} = - i [H, \rho] + \Gamma_0 \left(\sm \rho \sp - \frac{1}{2} (\sp \sm \rho + \rho \sp \sm)  \right),
\end{equation}
where $\rho$ is the system's density matrix, $\Gamma_0$ is the decay rate (the inverse of the natural lifetime) and $\sp$ and $\sm$ are raising and lowering operators with respect to $\ket{e}$ and $\ket{g}$. 

\begin{figure}
\begin{center}
\includegraphics[width=0.6\columnwidth]{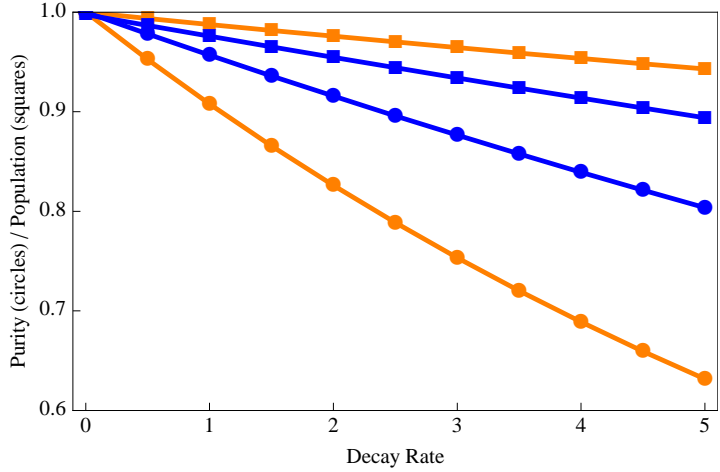}
\caption{Comparison of final population in the computational basis (squares) and final purity of the system's density matrix (circles) for the adiabatic gate (orange) and the dynamic gate (blue).  The decay rate is given in units of $\rm{ns}^{-1}$ and the system parameters are $J_1 = J_2 = 0.05~\rm{ps}^{-1}$ and $R=1.2$. Typical laser control parameters have been used, such that both gates achieve an entangling power very close to the maximal value of $e(U) \approx 0.22$. }
\label{decoherence_plot}
\end{center}
\end{figure}

We will explore two figures of merit: the amount of population returned to the desired computational basis states, and the purity of the density matrix, after application of the gate. For both gate types we have performed a full numerical simulation that for the dynamic gate includes the two (rectangular) laser $\pi$ pulses.
Fig. \ref{decoherence_plot} shows that the fast dynamic gate retains a higher purity even for fast decay rates. However, the adiabatic gate is more robust to loss of population from the computational basis.

\begin{figure}
\begin{center}
\includegraphics[width=0.6\columnwidth]{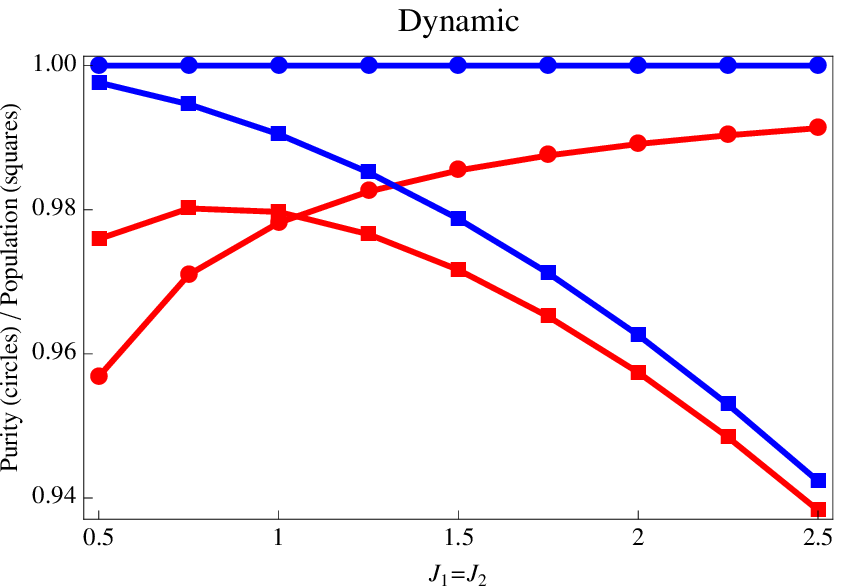}
\includegraphics[width=0.6\columnwidth]{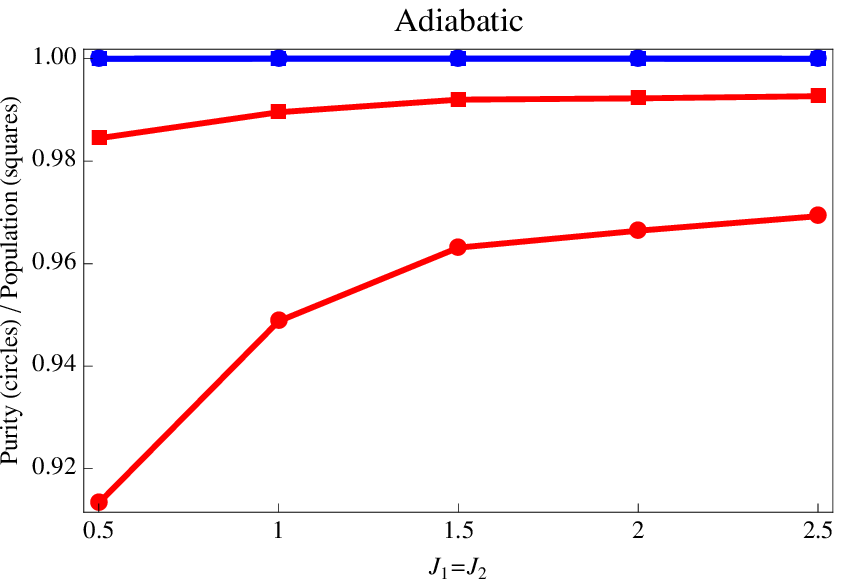}
\caption{Intrinsic gate errors (blue curves) and additional effects of decoherence (red curves): the final purity and the final population of the computional basis states are shown as a function of the coupling strength $J_1 = J_2$ (in units of $0.1~\rm{ps}^{-1}$). A decay rate of $1\,\rm{ns}^{-1}$,  $R=1.2$ and typical laser pulse parameters have been used. }
\label{decoherence_j1j2_plot}
\end{center}
\end{figure}

In Fig. \ref{decoherence_j1j2_plot}, we analyse the dependence of the two figures of merit on $J_1 = J_2$. For the dynamic gate, we keep $\Omega_0 = 5\,\rm{ps}^{-1} \approx 3\,\rm{meV}$ constant; this introduces an increasing intrinsic error as $t_\mathrm{rev}$ decreases and the transient regimes of excitation and de-excitation become more important. On the other hand, the purity improves as $t_\mathrm{rev}$ becomes shorter when radiative decay is included, as shown by the red curves. For the adiabatic gate, a pulse duration $\tau = 0.5\,\rm{ns}$ and coupling strength $\Omega = 0.066\,\rm{meV}$ are used, with the detuning adjusted in the range $\Delta = 0.16 -0.6\,\rm{meV}$ to give a  maximum entangling power for each of the $J_1=J_2$ points shown. As can be seen in Fig. \ref{decoherence_j1j2_plot},  the adiabaticity condition is very well satisfied and there is no population leakage in absence of radiative decay. However, decay events inevitably lead to some population leakage when included in the model. Unsurprisingly, a stronger $XY$ interaction improves performance for both purity and loss of population; this is in contrast to the dynamic scheme, for which the loss of population gets worse when the interaction strengths are larger. 

\section{Conclusion}

We have shown that it is possible for a central control to be a mediator of entanglement between two qubits, even if the coupling strengths between mediator and qubits are different. 
When a dynamic approach is taken, the coupling must be of $XY$ form -- but if an alternative adiabatic gate is performed {\it any} coupling form is permissible. Further, the gate is close to maximally entangling over a wide range of parameter space, in both cases.

The proposed protocol could immediately be tested in a range of experimental systems, include molecules with coupled electron and nuclear spins, and donors in silicon. Possible experiments would include the demonstration of remote entanglement between two centres that are not directly coupled, or the demonstration of a simple algorithm.

\section{Acknowledgments}

We would like to thank Simon Benjamin, Tony Harker and Dan Wheatley for valuable discussions, and Marcus Schaffry for a careful reading of the manuscript. We also thank the QIPIRC (No. GR/S82176/01) for support. EMG acknowledges support from the Marie Curie Early Stage Training network ÔQIPESTÕ (MEST-CT-2005-020505).  BWL acknowledges support from a Royal Society University Research Fellowship. AMS acknowledges support from EPSRC through the Basic Technologies project GR/S23506/01.

\end{document}